# Identification of hybrid node and link communities in complex networks

Dongxiao He[1,2], Di Jin[3], Weixiong Zhang[2,4,*]

[1] College of Computer Science and Technology, Jilin University, Changchun 130012, P. R. China
[2] Department of Computer Science and Engineering, Washington University, St. Louis, MO 63130, USA
[3] School of Computer Science and Technology, Tianjin University, Tianjin 300072, P. R. China
[4] Department of Genetics, Washington University School of Medicine, St. Louis, MO 63130, USA

Identification of communities in complex networks has become an effective means to analysis of complex systems. It has broad applications in diverse areas such as social science, engineering, biology and medicine. Finding communities of nodes and finding communities of links are two popular schemes for network structure analysis. These schemes, however, have inherent drawbacks and are often inadequate to properly capture complex organizational structures in real networks. We introduce a new scheme and effective approach for identifying complex network structures using a mixture of node and link communities, called hybrid node-link communities. A central piece of our approach is a probabilistic model that accommodates node, link and hybrid node-link communities. Our extensive experiments on various real-world networks, including a large protein-protein interaction network and a large semantic association network of commonly used words, illustrated that the scheme for hybrid communities is superior in revealing network characteristics. Moreover, the new approach outperformed the existing methods for finding node or link communities separately.

## 1. INTRODUCTION

Most complex systems in various fields, such as social networks in social science, the Internet in engineering, and signaling pathways in biology, can be formulated as networks where nodes represent entities (e.g., individuals in a social network) and links represent some relationship between nodes (e.g., co-worker relationship in a social network). Individual entities in a complex system seldom exist in isolation, but rather are often organized in groups to exert functions. For example, an organization is typically consisted of units of different but related functions that interconnect in particular structures to maximize the overall performance of the organization. In biology, a group of proteins in a cell interact to form an RNA polymerase for transcription of genes. Therefore, a critical step toward understanding complex systems is to uncover organizational or community structures in the networks [1]. Communities, also referred to as clusters or modules, are groups of nodes that share common properties or play similar roles [2]. A primary objective of community detection is to identify sets of nodes with common functions by using information of network topology.

Many methods for community identification have been proposed; the most popular ones belong to the scheme for detecting node community [1-8]. In this *node scheme*, communities are groups of nodes relatively densely connected within groups but sparsely connected across groups [4]. Indeed, many real networks carry structures that can form node communities [4-8].

In the conventional node scheme, a node belongs to only one community. However,

overlapping community structures are ubiquitous in real networks [9]. For example, an individual has a family and belongs to a group of co-workers, each of which has its own function and forms its own circle of influence. Forcing a node into one community will fail to honor possible multiple relationships and functions that a node may have, resulting in erroneous representation of the network structure [9].

To overcome this serious drawback, the link-community scheme has been proposed [10]. In this *link scheme*, links with a similar relational property form communities so that a node can inherit the community memberships of its adjacent links and, as a result, can naturally belong to multiple communities. Many real-world systems can be represented by link communities [10-14].

However, the link scheme typically generates a highly overlapping community structure even though a network has no overlapping structure at all [3]. Take the American college football network [4] as an example, which is to be discussed in the Experimental Analyses section. Under the link scheme, this network produced a highly overlapping community structure with 83 out of 115 nodes overlapped one another, despite that the football teams are organized in conferences that have no overlapping structure. This serious problem stems from the fact that the link scheme forces every link into a community while there are real networks that have links that do not fit into any community.

Many real-world systems have complex structures that are better characterized by a mixture of node and link communities. This suggests that a hybrid node-link community scheme, or *hybrid scheme* for short, will be more effective and robust in revealing and representing complex organizational structures than either the node or link community scheme. An illustrative example, shown in Fig. 1, is the network of 77 characters and their joint appearance in common scenes in Hugo's classic novel *Les Misérables*, which was complied by Knuth [15]. Here, nodes are characters and two nodes are connected if the two characters appear together in a scene. The node and link schemes produced distinctive community structures (Fig. 1A and 1B). Since a node was forced into one community in the node scheme, multiple community memberships were lost under this scheme. For example, Fantine was classified only into the pink community (the pink node in box to the left of Fig. 1A). In fact, Fantine and the seven blue nodes form another community (the clique of the seven blue nodes plus the pink node for Fantine in Fig. 1A), which is a small social group consisting of four Parisian students and their respective lovers. Therefore, the node scheme missed this important relationship between Fantine and this group she belongs to because it cannot properly characterize nodes with more than one role. This issue is exacerbated for the protagonist Valjean and his nemesis Javert (the other two pink nodes in box in Fig. 1A) who play more social roles than Fantine does and connect to ~48% of all the characters. The link scheme, on the other hand, may avoid such a problem by allowing nodes to exist in more than one community. However, it has its own drawbacks. For example, the link between Valjean and Bossuet was placed into the pink link community (the pink link connecting the two nodes in box in Fig. 1B) so that Bossuet was forced into this community. However, Bossuet does not appear together with the members of the pink community in any of the scenes except Valjean who belongs to not only the pink community but also four other communities. Thus, it is problematic to place Bossuet into the pink link community. A similar problem occurred with the link between Fantine and Thenardier as the latter does not appear together with the members of the pink community except Fantine. In sharp contrast, the hybrid scheme can provide elegant solutions to all these problems and correctly place multi-role characters into the right communities (Fig. 1C). Specifically, Fantine was put into both the blue link community and the pink node community, and Valjean and Javert were also correctly assigned to multiple communities, thereby fixing the problem of the node scheme. Moreover, the hybrid scheme did not force the link between Valjean and Bossuet and the link between Fantine and Thenardier into any community so that Bossuet (and Thenardier) was free from the pink community, fixing the problem of the link scheme.

However, it is challenging to detect hybrid node-link communities, which requires an accurate characterization of such structures. A viable approach is stochastic model which, instead of directly detecting communities, describes how such structures are generated in the first place. In this paper, we develop an effective approach based on stochastic model, named as NLC (Node-Link Communities), to identifying hidden hybrid node- and link-community structures in a network. The method can be run separately to find node, link or hybrid node-link communities as so desired.

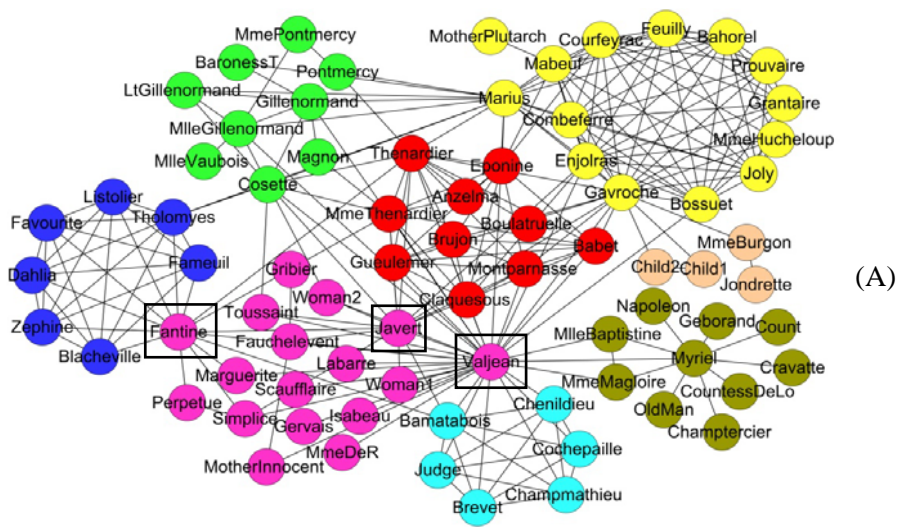

(A)

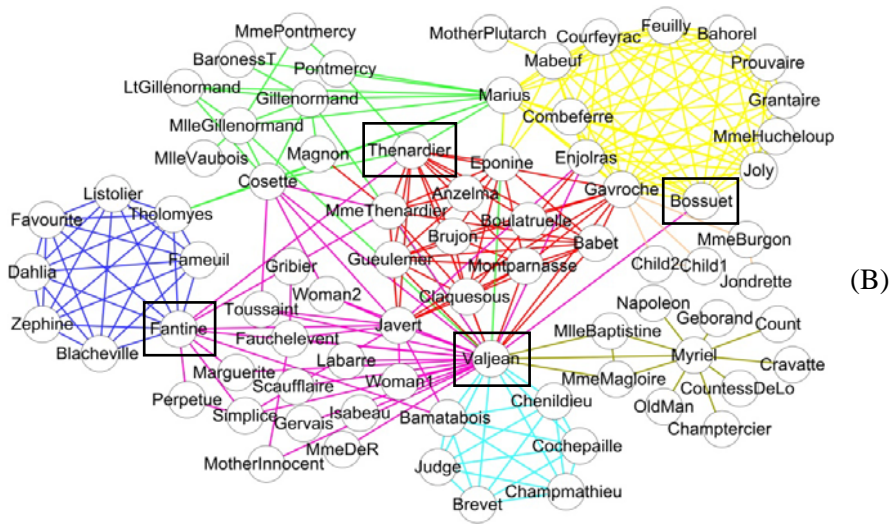

(B)

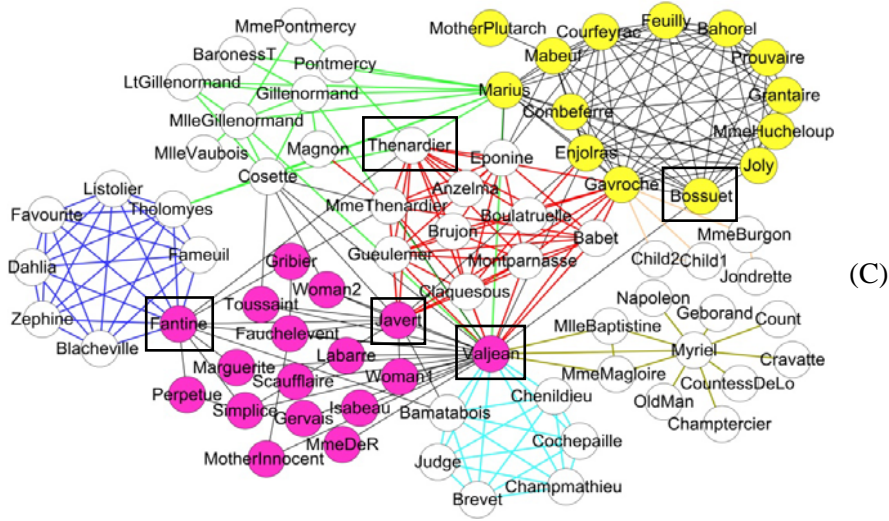

(C)

**Fig. 1.** Community structures of the co-appearance network of characters in *Les Misérables* from (**A**) the node scheme, (**B**) the link scheme and (**C**) the hybrid scheme. Here, node or link communities are colored in nodes or links respectively, and uncolored nodes and black links represent background.

## 2. THE METHOD

A central piece of our method is a probabilistic model to fit a given network. We are particularly interested in such a model that can accommodate both node and link communities. For clarity, in the subsequent sections we first describe the model and estimation of its parameters, and then consider inferring the hybrid node-link community structure from the model constructed.

### 2.1. Stochastic Model of Node and Link Communities

We first discuss the model and then consider learning its parameters by a method for maximizing a likelihood function to fit the model to a given network.

*2.1.1. The model*

Our model consists of a set of probabilistic node and link communities that best fit a given network. In this model, a node (or a link) has a probabilistic membership in a node (or link) community, and the nodes (or links) that have high probabilities of a common membership form a probabilistic node (or link) community. In this formulation, we only need to focus on expected memberships. Specifically, given a network with $n$ nodes, the model $G$ can be specified by a set of parameters $\{d_{i1}, d_{i2},…,d_{ic}\}$ for each node $i$, for $i=1,2,…,n$, and a total of $c$ communities, where $d_{ik}$ is proportional to the expected membership of node $i$ in the $k$-th probabilistic community $G_k$. That is, if $G_k$ is a node community, $d_{ik}$ is the expected node degree of $i$ in $G_k$, otherwise (i.e., $G_k$ is a link community), $d_{ik}$ is the expected number of links belonging to $G_k$ that node $i$ connects to; $d_{ik}$'s in both cases are equivalent.

It is critical to note that a community in the model has no further subdivision and can be regarded as a random graph with no community structure. Therefore, a random-graph null model (namely null model of modularity) [5], which describes a random graph with a sequence of node degrees and with edges drawn at random among the nodes, can be adopted to characterize each of the communities. Following this null model, the expected number of links (or expected link weight) between nodes $i$ and $j$ in $G_k$ can be evaluated as

$$\hat{w}_{ij}^k = \frac{d_{ik}d_{jk}}{\sum_s d_{sk}}. \tag{1}$$

The expected number of links between nodes *i* and *j* in the given network can then be written as

$$\hat{w}_{ij} = \sum_k \hat{w}_{ij}^k = \sum_k \frac{d_{ik}d_{jk}}{\sum_s d_{sk}}. \tag{2}$$

Note that multiple links between two nodes and self-edges are allowed here, which is typical for random graph models for simplicity [13, 16, 17]. The property of multiple links makes the model applicable to some weighted networks.

Intuitively, if node *i* in the model has a large membership in a node community, it should have a high probability to connect with other nodes in that community, and consequently, all nodes with large memberships in a common community tend to be densely connected to form the node community. Likewise, all nodes with high memberships in a link community, as they have large numbers of adjacent links of a common type, tend to be highly connected through the same type of links to form the link community.

*2.1.2. Parameter learning*

Our next step is to learn the parameters of the model to describe the community structure implied by $d_{ik}$'s. This is done by maximizing a likelihood function that the given network was generated from the model. Because the number of links between two nodes, $w_{ij}$, is Poisson distributed with its expectation $\hat{w}_{ij}$ [13, 17], the probability for generating a graph *G* with adjacency matrix $A = (w_{ij})_{n \times n}$ following (2) is

$$P(G|d) = \prod_{i,j} \frac{\left(\sum_k \frac{d_{ik}d_{jk}}{\sum_s d_{sk}}\right)^{w_{ij}}}{w_{ij}!} \exp\left(-\sum_k \frac{d_{ik}d_{jk}}{\sum_s d_{sk}}\right). \tag{3}$$

The best fit between the expected graph following (2) and the given network can be achieved by maximizing the likelihood function in (3). To be effective, the maximization is typically done with the logarithm of the likelihood, which has no effect on the position of the maximum. Applying logarithm to (3), rearrange, and dropping additive and multiplicative constraints, we derive the log likelihood

$$L = \sum_{ij} w_{ij} \log\left(\sum_k \frac{d_{ik}d_{jk}}{\sum_s d_{sk}}\right) - \sum_{ijk}\left(\frac{d_{ik}d_{jk}}{\sum_s d_{sk}}\right). \tag{4}$$

Direct maximization of (4) seems to be nontrivial. Here we adopt an expectation-maximization (EM) algorithm [18]. By applying Jensen's inequality to (4), we have

$$L \geq \bar{L} = \sum_{ijk}\left(w_{ij}q_{ij,k}\log\frac{d_{ik}d_{jk}/\sum_s d_{sk}}{q_{ij,k}} - \frac{d_{ik}d_{jk}}{\sum_s d_{sk}}\right), \tag{5}$$

where the probabilities $q_{ij,k}$ can be freely chosen, provided that they satisfy $\sum_k q_{ij,k} = 1$. The

exact equality is reached when

$$q_{ij,k} = \frac{d_{ik}d_{jk}/\sum_s d_{sk}}{\sum_r (d_{ir}d_{jr}/\sum_s d_{sr})}. \tag{6}$$

Thus, the double maximization of the new function $\bar{L}$ with respect to the model parameters $d_{ik}$ and the probabilities $q_{ij,k}$ is equivalent to maximizing the original log-likelihood $L$ with respect to the model parameters alone. Given the optimal model parameters $d_{ik}$, the optimal probabilities $q_{ij,k}$ satisfies (6) since these values make the inequality in (5) an equality. Given the optimal probabilities $q_{ij,k}$, the optimal values of model parameters $d_{ik}$ can be derived by maximizing $\bar{L}$. Then by differentiating $\bar{L}$ in (5), we obtain the optimal values of $d_{ik}$

$$d_{ik} = \sum_j w_{ij} q_{ij,k}. \tag{7}$$

Maximizing the log-likelihood $L$ is to simultaneously solve (6) and (7), which can be done iteratively by choosing a set of initial values and alternating between the two equations. This approach, known as the expectation-maximization algorithm, can monotonically converge to a local minimum of the log-likelihood function.

Now consider the complexity of the learning algorithm. Note that $q_{ij,k}$ is defined for a node pair $(i, j)$ connected by an edge in the network so that $A_{ij} = 1$. Thus the time to evaluate (6) once is $O(mc)$, where $m$ is the number of edges and $c$ the number of communities. The time to calculate (7) once is $O(mc)$ as well because we only need to consider the observed edges. Consequently, the time complexity of the learning algorithm is $O(Tmc)$, where $T$ is the number of iteration executed.

**2.2. Inferring Hybrid Community Structure**

Even with a model of node and link communities constructed for a given network, it is not straightforward to infer community structures. This entails inferring the nodes or links, respectively, in a node or link community, and determining the type (i.e., node or link) of each of the communities. For clarity, we consider these two issues separately.

*2.2.1. Inferring community structure given the types of communities*

Determining the structure of a community amounts to determining its members. Assume that the type of each of the communities is known. We first define two sets of variables: $S_i^k$ represents the probability or probabilistic membership that node $i$ belongs to the $k$-th community $G_k$, and $R_{ij}^k$ denotes the probability that a link $<i, j>$ belongs to $G_k$. Then, $S_i^k$ can be evaluated as

$$S_i^k = \frac{d_{ik}}{\sum_r d_{ir}}, \tag{8}$$

and $R_{ij}^k$ can be written as

$$R_{ij}^k = \frac{\hat{w}_{ij}^k}{\hat{w}_{ij}} = \frac{d_{ik}d_{jk}/\sum_s d_{sk}}{\sum_r (d_{ir}d_{jr}/\sum_s d_{sr})}, \tag{9}$$

The probabilistic memberships of communities are used to infer deterministic memberships of communities, thus forming deterministic communities defined as $\{H_1,$

$H_2,…,H_c$}. If $H_k$ is a node community, it will consist of all nodes $i$ satisfying $arg\max_r\{S_i^r \mid r = 1,2,…,c\} = k$; if $H_k$ is a link community, it will contain all links $<i, j>$ satisfying $arg\max_r\{R_{ij}^r \mid r = 1,2,…,c\} = k$.

*2.2.2. Determining the types of communities*

Determining the type of each of the communities separately seems to be nontrivial, and may not necessarily give rise to a global optimality for the whole network either. Here we adopt a global method for this problem. Since there are $c$ communities, each of which can be either a node or link community, there are a total of $2^c$ possible combinations of hybrid node-link communities. In order to identify the best among these combinations, we need a quality metric to measure the quality of a candidate combination of communities.

The quality measure of community structure that we adopt is the revised map equation (see Appendix A for detailed description) [7, 19], which is based on the principle of minimum description length (MDL) [20] and is suitable for overlapping communities. Under this measure, the shorter the MDL of an overall community structure, the better the structure is. For clarity, we use $V_k$ or $E_k$ to explicitly indicate that the $k$-th deterministic community $H_k$ is a node or link community, respectively. Assume that $H = \{H_k \mid k = 1,2,…,c\}$ is a candidate hybrid node-link community structure, where $H_k$ is either $V_k$ or $E_k$. Let $L(H)$ be the value of MDL for $H$. A straightforward way to finding the best hybrid node-link community structure is to enumerate all possible combinations for $H$ to find the one with the minimum value of MDL. This exhaustive search may become computationally expensive for large networks. Here we offer an alternative, an effective heuristic, to this search problem, which takes the following steps.

**S1** Initialize a candidate hybrid community structure $H$: for community $k$, randomly assign either $V_k$ or $E_k$ to $H_k$;

**S2** Update $H$: for community $k$, swap the current $H_k$ to the other community ($V_k$ or $E_k$) if doing so reduces $L(H)$;

**S3** Repeat S2 until $L(H)$ cannot be reduced further, or the maximal number of iterations has been executed.

According to [19], the time to evaluate the MDL-value of a community structure $H$ once is $O(lmo^2)$, where $l$ is the number of steps of a random walk in map equation, $m$ the number of edges, and $o$ the average number of communities that each node belongs to. Moreover, $L(H)$ needs to be evaluated $K \times c$ times in the above heuristic algorithm, where $K$ is the maximal number of iterations and $c$ the number of communities. Thus, the time to determine the types of all communities in the algorithm is $O(Kclmo^2)$. Recall that the time for learning the model parameters in Sec. 2.1.2 is $O(Tmc)$. Thus, the time complexity of the NLC method is $O((mc)(T+Klo^2))$. Notice that, $T$, $K$ and $l$ are often taken as constants, and $o$ is usually small and bounded by a constant. Therefore, the total time complexity of NLC can be considered as $O(mc)$.

## 2.3. A Working Example of NLC

Here we demonstrate the working of our method NLC with an example. The given network is shown in Fig. 2A. Given the model parameters $d_{ik}$ (see Table 1 and discussion below), we can form the expected graphs of all the communities of the observed network (Fig. 2B, 2C and 2D) according to (1). Further, we can form the expected graph of the whole graph $G$ (Fig. 2E) according to (2), which is an ensemble of the expected graphs of all its communities.

However, since the model parameters are unknown *a priori*, we fit network and its expected graph by optimizing (3) to learn the best $d_{ik}$ (Table 1). Subsequently, we infer all the node and link communities according to (8) and (9), and identify the final network community structure (Fig. 2F) based on the principle of minimum description length.

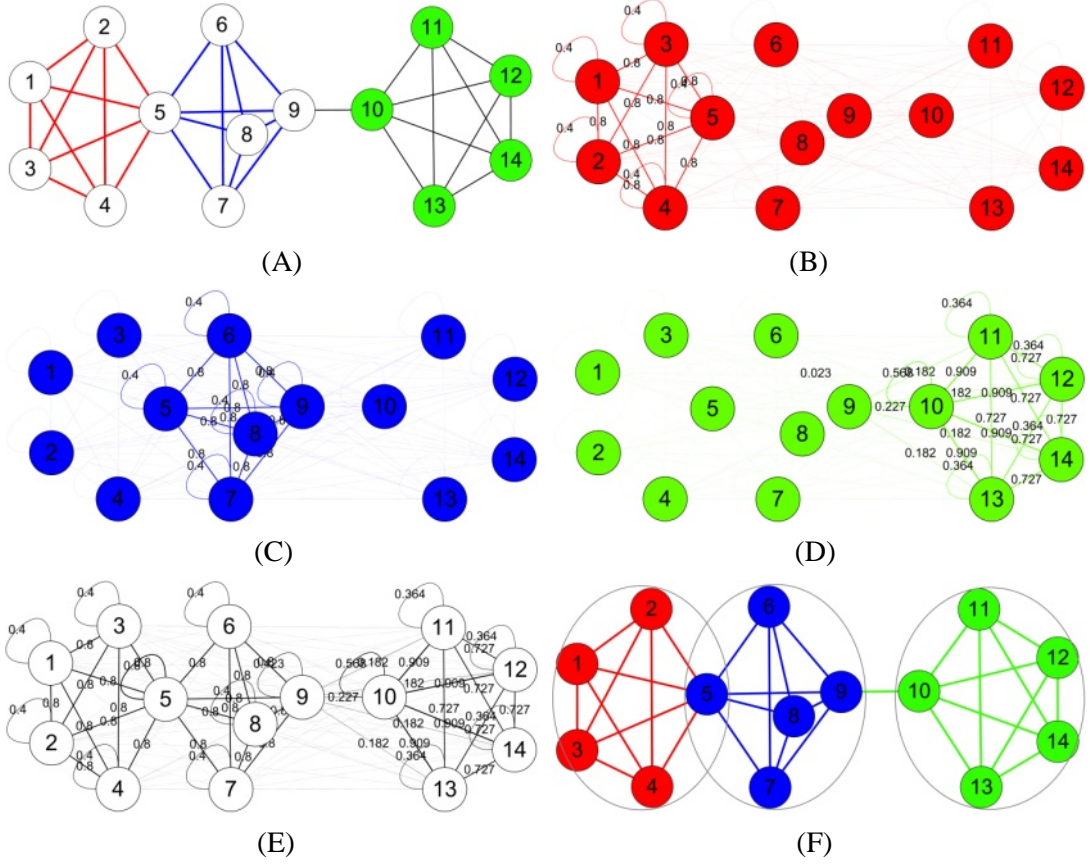

**Fig. 2.** An illustration of NLC for identifying hybrid node-link community structures. **(A)** A given network $G$ with two link communities (in red and blue) and one node community (in green). **(B)**, **(C)** and **(D)** The expected graph of the red, blue and green community. Note that the width of a link corresponds to its expected values, and the values smaller than 1.0e-3 are omitted. **(E)** The expected graph of $G$, which is an ensemble of the expected graphs of the red, blue and green communities. **(F)** The final inferred node and link communities colored in nodes or links respectively, and noted by three cycles.

**Table 1** The learned model parameters $d_{ik}$'s

| $d_{ik}$ | $i$=1 | $i$=2 | $i$=3 | $i$=4 | $i$=5 | $i$=6 | $i$=7 | $i$=8 | $i$=9 | $i$=10 | $i$=11 | $i$=12 | $i$=13 | $i$=14 |
|---|---|---|---|---|---|---|---|---|---|---|---|---|---|---|
| $k$=1 | 4.73e-124 | 3.04e-123 | 1.92e-123 | 1.02e-123 | 2.84e-17 | 3.89e-15 | 4.30e-15 | 3.84e-15 | 0.999991 | 4.999991 | 4 | 4 | 4 | 4 |
| $k$=2 | 3.999987 | 3.999987 | 3.999986 | 3.999987 | 3.999946 | 3.24e-20 | 5.36e-19 | 8.14e-20 | 1.15e-14 | 4.14e-103 | 0 | 0 | 0 | 0 |
| $k$=3 | 1.35e-05 | 1.35e-05 | 1.35e-05 | 1.35e-05 | 4.000054 | 4 | 4 | 4 | 4.000009 | 9.04e-06 | 6.50e-87 | 8.66e-86 | 8.05e-86 | 4.21e-86 |

## 3. EXPERIMENTAL ANALYSES

We performed two sets of experiment to demonstrate the favorable features of the new scheme of hybrid communities over the existing schemes of single type of communities and to show the superior performance of our method NLC. In our experiments, since our model needs the number of communities $c$ as a parameter, we used the MDL as a yardstick to search for the target community structure by iterating over possible values of $c$.

## 3.1. Comparison of the Three Community Schemes

The NLC method supports the three schemes for finding hybrid community structures as well as node and link communities separately. We thus applied it to identify the best network structures for each of these schemes. The comparison was done on three real network problems.

### 3.1.1. Zachary's karate club

The Zachary's "karate club" network [21] has become a *de facto* testbed for community detection algorithms. Fig. 3 shows the community structures from the three schemes compared. The result from the node scheme has three disjoint node communities (Fig. 3A). In this partition, node 1 (the instructor, the red node in box) was exclusively assigned to the red community, even though it is also connected to all nodes (except one) in the purple community, showing its important role in the purple community. In comparison, the hybrid (and link) scheme (Fig. 3B) correctly placed node 1 (square node in box) in the purple and red link communities. Moreover, the MDL of the result of the hybrid (and link) scheme is 4.2966, which is smaller than that for node partition (4.3563).

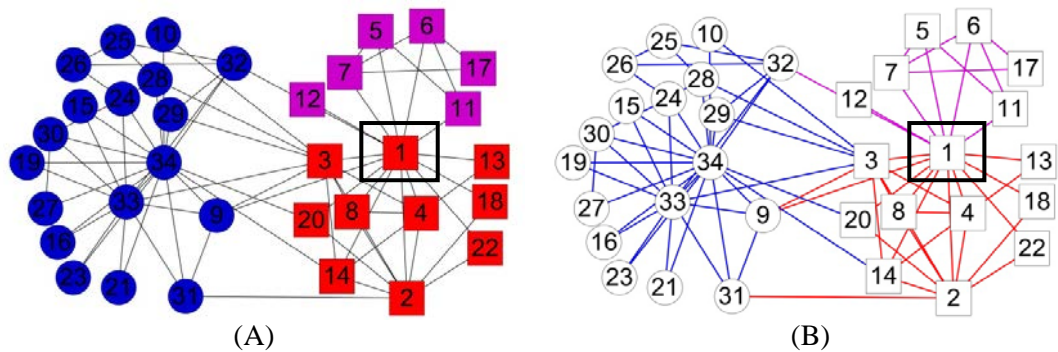

(A)                   (B)

**Fig. 3.** Communities of the "karate club" network obtained by **(A)** the node scheme and **(B)** the hybrid (and link) scheme. The nodes in circle and square represent the two communities as originally reported: the club administrator's faction in circles and the instructor's faction in squares. Node or link communities from our model are colored in nodes or links, respectively.

To further evaluate the quality of the results from the three schemes, we compared the MDLs of the structures from these schemes with the community number $c$ varied. As shown in Fig. 4, the results from the hybrid scheme have shorter MDLs than the other two schemes except when $c$ equals to 3 at which the hybrid and link schemes produce the same network structure. The result in Fig. 4 also suggested that there should be 3 communities, whereas the reported "actual" number was two. In fact, the instructor's faction (square nodes) contains two evidently overlapping subgroups that were connected only through the instructor (node 1, Fig. 3B). Thus, it is more reasonable to split the instructor's faction into two.

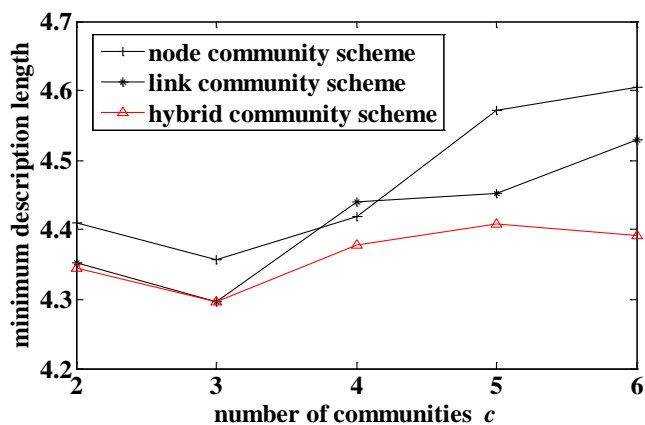

**Fig. 4.** The minimum description lengths for the results from the three schemes with varying number of communities $c$ on the "karate club" network. As shown, the hybrid scheme produced structures with the smallest MDLs and the best structure has three communities.

*3.1.2. American college football network*

In the American college football network [4], the nodes represent football teams and a link represents a game played by two teams during the football season in year 2000. The teams are divided into "conferences", which form actual communities. Teams in the same conference play more often than teams not in the same conference. A team plays an average of approximately 7 intra- and 4 inter-conference games in the season. This suggests that the network processes typical characteristics of node communities. As expected, the hybrid scheme discovered a node-community structure for this network which is shown in Fig. 5A. In contrast, the link scheme produced a highly overlapping community structure with 83 out of all 115 nodes overlapped shown in Fig. 5B, revealing a serious drawback of this scheme. Since the reported community structure is known, we evaluated the results from the hybrid and link schemes against the reported network structure by the Extended Normalized Mutual Information (ENMI) index [22]. The hybrid scheme scored ENMI=0.8035 while the link scheme scored ENMI=0.3604, evidently showing that the former outperformed the latter significantly. Furthermore, we also compared the community structures from hybrid (node) scheme and link scheme as well as the reported community structure using the MDL quality metric. The MDL for the hybrid scheme (5.4487) is smaller than that for the link scheme (6.1125). Surprisingly, the MDL for the hybrid scheme is also smaller than that of the reported structure (5.6772). This maybe due to two factors. First, the independent teams that do not belong to any conference are grouped into a separate but subjective "conference" in the reported community structure even though these independent teams do not play more often among themselves than with other teams. Second, our hybrid, data-driven community discovery scheme is able to more faithfully detect community structures underlying the data of overall games played than the reported result.

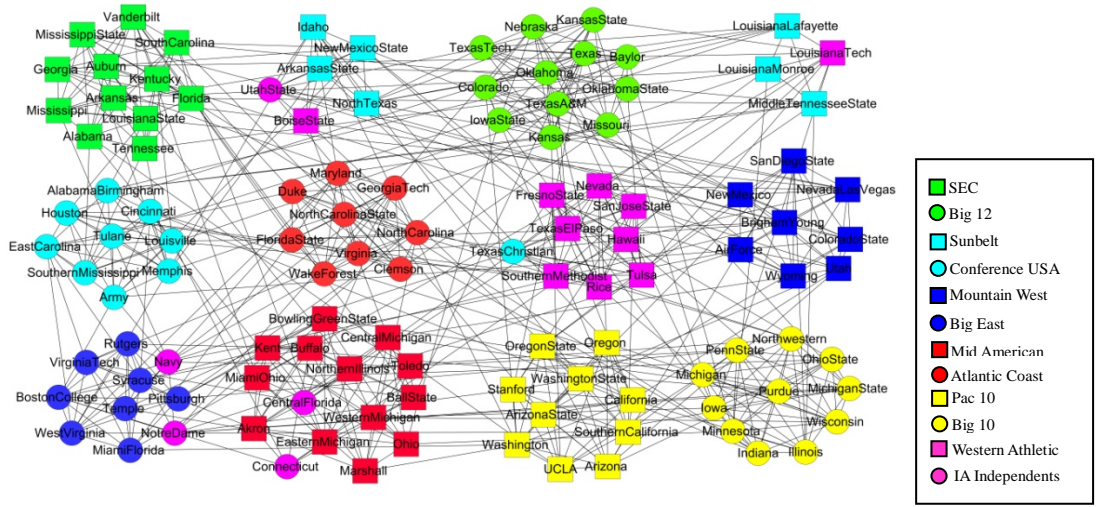

(A)

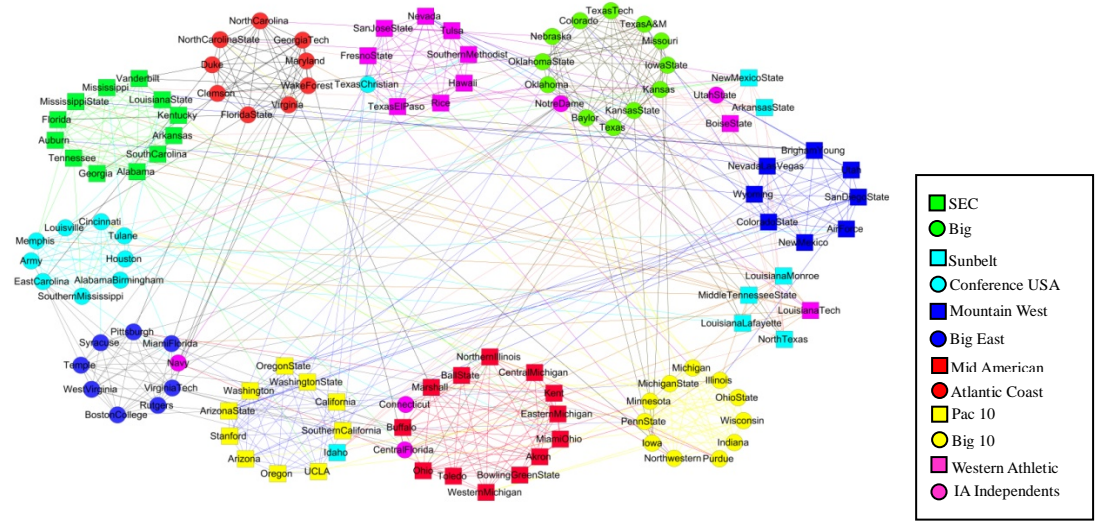

(B)

**Fig. 5.** Community structures of American college football network obtained by **(A)** hybrid and node-community schemes and **(B)** the link-community scheme. Nodes in the network represent teams and links represent games between teams. Here, the 12 different combinations of node shape and node color represent the actual "conferences". The clusters of nodes in space denote node communities obtained by our model in (A), and the colored links denote link communities from our mode in (B).

The MDL was used to evaluate the community structures obtained by the three schemes with varying number of communities. The detailed result is shown in Fig. 6. As this network has typical characteristics of node communities, the hybrid scheme always produced the same results as the node scheme, and the MDLs from the hybrid (and node) scheme are always smaller than that of the link scheme. The best network structure was found by the hybrid scheme with 12 node communities, which perfectly matched the actual number of conferences.

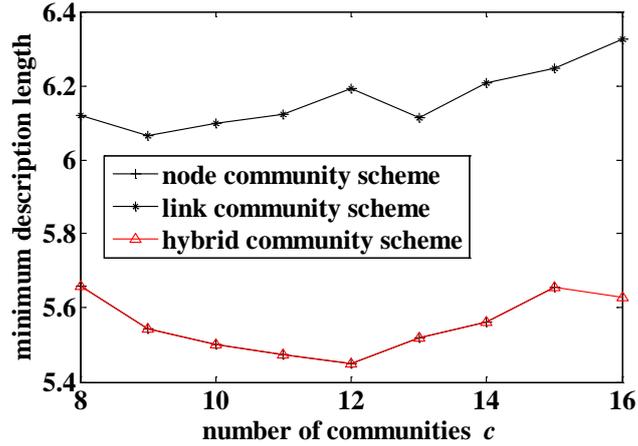

**Fig. 6.** The minimum description lengths for results from the 3 schemes with varying number $c$ of communities on the American college football network.

### 3.1.3. Les Misérables

The three distinct community structures for the three schemes are in Fig. 1. As discussed in Introduction, the hybrid scheme can overcome the shortcomings of the node and link schemes. Furthermore, the MDL of the result from the hybrid scheme (4.6783) is less than that from the node scheme (4.7528) and link scheme (4.7259). Similar to the two early network problems, the results from the hybrid scheme on this co-appearance network have shorter MDLs than the node and link schemes with all values of community number $c$ evaluated, which is shown in Fig. 7. The shortest description length appeared with 8 communities (Fig. 1).

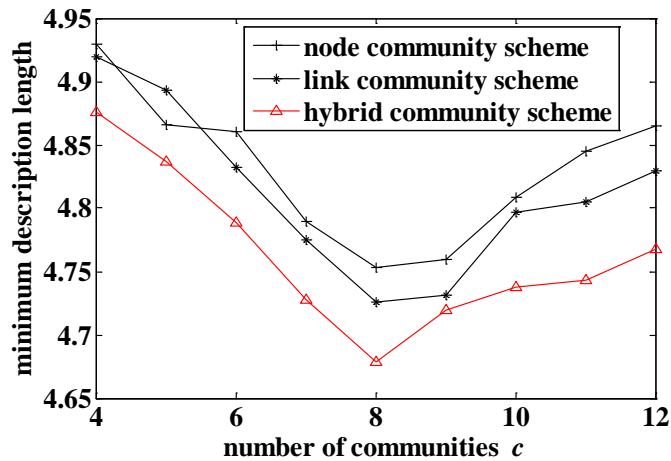

**Fig. 7.** The minimum description lengths for results from the 3 schemes with varying number $c$ of communities on *Les Misérables*.

### 3.2. Comparison with the Existing Methods

We evaluated the performance of NLC, designed primarily for finding hybrid node-link communities, along with several well-established methods for finding node communities or link communities on eight widely used real networks (Table 2 and Table 3). The methods compared include: the Louvain method [6] which is regarded as one of the best for node partitioning [1], LC (Link Community) [10] which is the most well-known method for link-community finding, and CPM (Clique Percolation Method) [9] which is the most prominent algorithm for overlapping community detection. We also included in the comparison two model-based methods proposed by Newman *et al*, i.e., NModel for node communities [17] and LModel for link communities [13].

Measured by the MDL, the new method NLC has the best performance on all networks except on Karate where LModel performed slightly better (Table 2). The superior performance of NLC is attributed to its flexibility and robustness in forming hybrid node-link communities. For example, on the co-appearance network of *Les Misérables*, NLC identifies a hybrid community structure with 6 link and 2 node communities with a MDL of 4.6783 (Fig. 1C), which is better than the node partition from NModel (5.2496) and the link partition from LModel (4.7784).

**Table 2**. Comparison of the minimum description lengths of the results from the new NLC method and five existing algorithms on eight real-world networks obtained from Newman's website [23]. Here, *n* is the number of nodes and *m* the number of links, and 'node', 'link', 'overlap', and 'hybrid' denote node, link, overlapping, and hybrid communities, respectively. The shorter the MDL of an overall community structure, the better the structure is. The best MDLs for these networks are underlined.

| Datasets | n | m | Louvain (node) | LC (link) | CPM (overlap) | NModel (node) | LModel (link) | NLC (hybrid) |
|---|---|---|---|---|---|---|---|---|
| Zachary's karate club | 34 | 78 | 4.3359 | 5.2502 | 5.8552 | 5.2365 | **4.2961** | *4.2966* |
| American college football | 115 | 613 | 5.4982 | 7.7749 | 5.5376 | 5.7852 | 6.1475 | **5.4487** |
| Les Miserables | 77 | 254 | 4.7632 | 5.4371 | 5.4126 | 5.2496 | 4.7784 | **4.6783** |
| Dolphin social network | 62 | 160 | 4.8859 | 6.6531 | 6.0031 | 5.4620 | 4.9687 | **4.8247** |
| Political books | 105 | 441 | 5.5836 | 7.6100 | 5.7855 | 5.5654 | 5.5452 | **5.5425** |
| Jazz musicians collaborations | 198 | 2,742 | 6.8745 | 8.9557 | 7.3312 | 6.8659 | 6.8737 | **6.8529** |
| C. Elegans neural network | 297 | 2,148 | 7.6309 | 11.2642 | 8.0112 | 7.9074 | 7.5941 | **7.5627** |
| E-mail network URV | 1,133 | 5,451 | 8.5428 | 12.2934 | 9.5409 | 8.6736 | 9.0147 | **8.4780** |

To further assess the qualities of the results from different methods, we used conductance [24], which is a well-known criterion for measuring the quality of a community. However, the original conductance is for assessing a single community. We extended it to a community structure by computing the weighted average of the conductance values of all communities to be measured.

The conductance of a community $S$ can be simply thought of as the ratio between the number of edges within the community and that across it. Formally, the conductance of a community $S$ is

$$\phi(S) = \frac{\varphi(S)}{\min(Vol(S),\ Vol(V \setminus S))}, \quad (10)$$

where $\varphi(S) = |\{(i, j) : i \in S, j \notin S\}|$, $Vol(S) = \sum_{i \in S} d_i$, and $d_i$ is the degree of node *i*. Thus, the lower the conductance of a community, the better it is. Consequently, the weighted average conductance (WAC) can be defined as

$$WAC = \frac{1}{\sum_{k=1}^{c} N(C_k)} \sum_{k=1}^{c} N(C_k)\phi(C_k), \qquad (11)$$

where $c$ is the number of communities, $C_k$ is the $k$-th community and $N(C_k)$ is the number of vertices in $C_k$.

Measured by the WAC, our NLC has the best performance on all the eight networks except on Jazz where CPM and NModel performed slightly better (Table 3). This set of results, which is independent of MDL index, can further illustrate the favorable property of the new scheme of hybrid communities over the existing schemes of single type of communities, and show the superior performance of our method NLC.

**Table 3**. Comparison of the weighted average conductances of the results from our method NLC and five existing algorithms on eight real-world networks. The smaller the WAC of an overall community structure, the better the structure is. The best WACs for these networks are underlined.

| Datasets | Louvain (node) | LC (link) | CPM (overlap) | NModel (node) | LModel (link) | NLC (hybrid) |
|---|---|---|---|---|---|---|
| Zachary's karate club | 0.4141 | 0.5917 | 0.8365 | 0.5602 | **0.2066** | **0.2066** |
| American college football | **0.3110** | 0.7155 | 0.3531 | 0.3863 | 0.4706 | **0.3110** |
| Les Miserables | 0.3343 | 0.6197 | 0.5027 | 0.4019 | 0.4233 | **0.3084** |
| Dolphin social network | 0.3128 | 0.7187 | 0.7161 | 0.4434 | 0.3207 | **0.3105** |
| Political books | 0.2985 | 0.6873 | 0.3412 | 0.1024 | 0.1416 | **0.0970** |
| Jazz musicians collaborations | 0.3344 | 0.7151 | **0.0716** | *0.1932* | 0.2170 | *0.1944* |
| C. Elegans neural network | 0.4816 | 0.8519 | 0.3406 | 0.3422 | 0.2467 | **0.2259** |
| E-mail network URV | 0.4298 | 0.8727 | 0.7220 | 0.3919 | 0.4680 | **0.3531** |

## 4. APPLICATIONS TO LARGE NETWORKS

We applied the hybrid scheme and the NLC algorithm to help elucidate structures of a large protein-protein interaction network in biological science and reveal hidden associations among words commonly used. The domain specific results, such as protein-protein interactions and their biological implication, will be reported elsewhere. Here, we discuss the results on network structures identified as a whole using an objective quality measure and in comparison with some of the existing methods.

In order to obtain an objective quality assessment beyond a network structural measure, such as the MDL, we adopted domain knowledge specific to these applications to assess the quality of the results.

Furthermore, the study was performed in comparison with two well-known methods that can discover overlapping communities and are applicable to large networks. The first is CPM [9], which is the most prominent algorithm for detection of communities with overlapping structures. Of note, CPM may treat some nodes of a network as background and not classify them into any community. To set a base for comparison, we took the subgraph processed by CPM as the targeted network, and used the number of communities attained by CPM as the number of communities for our model. Note that this comparison was done on a subgraph rather than on the whole network. The second comparison was carried out on the whole networks and between NLC and LC [10], which is best known for link-community finding and is able to handle large networks. As the two networks considered here are much larger than those listed in Table 2, it is not practical to determine the number of communities *c* by searching for the best structure among all candidates with different *c*. Thus, we adopted a simple partitioning strategy in NLC by repeatedly bipartitioning a (sub)network using our model until the likelihood function could not be further improved. This strategy makes NLC a hierarchical clustering algorithm similar to the LC method, making the two methods more comparable for performance evaluation.

**4.1. Protein-Protein Interaction Network**

The first large network considered is the protein-protein interaction (PPI) network of budding yeast *Saccharomyces cerevisiae* [25]. It contains 2,640 nodes (proteins) and 6,600 links (physical interactions between pairs of proteins).

We used the Gene Ontology (GO) terms [26], the most elaborate gene function annotations, as domain metadata for quality assessment. The GO terms include information of functions and cellular locations of a gene and biological pathways that a gene may be involved. The biological significance of a community of genes (nodes) can be measured by the GO terms enriched in the genes in the community. Enrichment of GO terms is evaluated by a hyper-geometric test [27], providing a GO term a *p*-value to quantify the significance of the term. To measure the biological significance of a community structure, we used as a quality metric the average number of significantly enriched GO terms with *p*-values not exceeding a threshold. The larger this average number of significant GO terms, the more biologically significant a community structure is.

The NLC method identified PPI community structures with many more significant GO terms than CPM under all 10 different *p*-value thresholds tested (Fig. 8A), showing the superior performance of NLC over CPM. Note that NLC was run on the subgraph that was sifted through by CPM, making the comparison biased in favor of CPM. Furthermore, NLC outperformed LC as well using the same quality assessment (Fig. 8B).

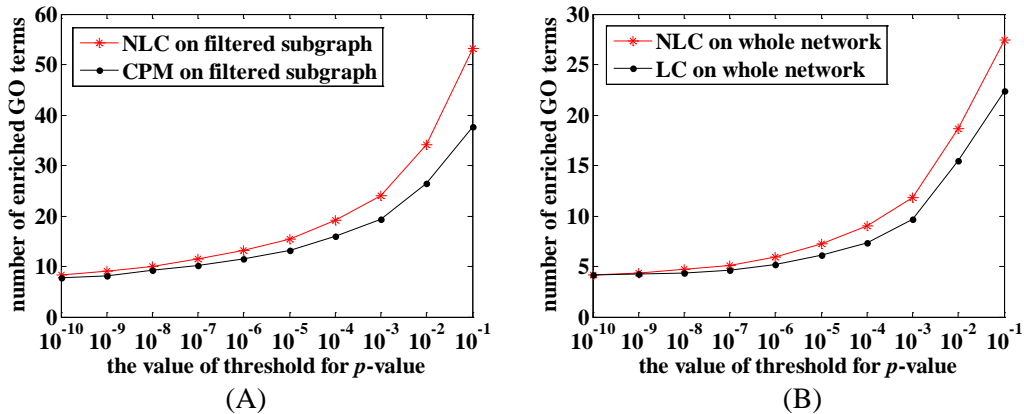

**Fig. 8.** Comparison of NLC and CPM (**A**) and NLC and LC (**B**) on a large budding yeast PPI network in terms of the number of enriched GO terms that are statistically significant with hyper-geometric *p*-values below a threshold.

**4.2. Word Association Network**

The second large network deals with words and the associations among words that people typically intended to use. The network, with 5,017 nodes (words) and 29,148 links

(association between pairs of words), is constructed from the University of South Florida Free Association Norms data set [28] in the manner of [9].

We adopted WordNet, which is an online lexical reference database [29], as the domain metadata for quality assessment of community structures. In WordNet, words are organized in sets of cognitive synonyms, known as Synsets, each of which represents one lexical concept. In our analysis, we considered two words to be semantically related or similar when they belong to the same Synset. To assess the quality of a community structure, we computed the enrichment of similarity between a pair of nodes [10]

$$\text{Enrichment} = \frac{\langle \mu(i,j) \rangle_{\text{all } i,j \text{ within same community}}}{\langle \mu(i,j) \rangle_{\text{all pairs } i,j}}, \tag{12}$$

where $\mu(i,j)=1$, if words $i$ and $j$ belong to the same Synset, or 0, otherwise. The larger the enrichment, the better a community structure is.

We first compared NLC with CPM, following the same comparison scheme as in the PPI network analysis. The enrichments of the results from NLC and CPM were 32.2179 and 28.0801 respectively, showing the superiority of NLC over CPM even though the comparison was in favor of the latter as the subgraph chosen by CPM was analyzed. In the second comparison against LC, the result from NLC had an enrichment value of 72.9202, which was greater than that of 71.5827 from LC. Therefore, NLC is more effective than the two popular existing methods in revealing semantic associations among words in the large word network.

## 5. DISCUSSIONS

This is the first time that a hybrid node-link community scheme has been proposed for characterizing complex network structures and an effective model and algorithm were developed for finding such hybrid communities. The hybrid scheme is able to overcome the inherent drawbacks of the node and link schemes, such as inability to support multiple roles that a node may play or forcing nodes to have unsupported relationships, which restrained the applicability of the node and link schemes. The analyses on several real networks, including a large protein-protein interaction network and a large word association network, demonstrated the superb properties of the hybrid scheme in revealing subtle and intricate network structures in real networks. The new NLC method, whose software is available from the authors, can be used separately to find node, link and hybrid node-link communities.

Stochastic models have been proposed, separately, for node communities [17, 30-33] and link communities [13]. However, they fail to model the two types of communities together. Here, we developed a unified model of node and link communities. Different from the existing models [13, 17, 30-33] that extend the classic stochastic blockmodel [34], our new model generalizes the null model of modularity [5] to incorporate the ability of describing mixed communities which the original null model does not possess. At the center of our method using the model is an expected node degree function, which is optimized to fit the node degrees of a given network by an expectation-maximization algorithm. The detailed comparisons of our model with some existing ones are given in the following.

Model-based method is a promising class of techniques for identifying network communities, which have been under active research and development [35]. Several similar methods have been proposed, which are based on stochastic modeling and the expectation-maximization algorithm or nonnegative matrix factorization for optimization; however, most of the existing methods have focused on detection of node communities [17, 30-33].

One exception to node community detection using stochastic modeling is due to Ball, Karrer & Newman [13], which considers the detection of link communities. Meanwhile, as the authors pointed out, their model can be extended to detect node community by introducing a block matrix. Note that, Ball's model is just LModel in the part of experimental comparison. Although Ball's model and our model presented here seemed to be similar, they have several key differences. In stochastic models for node communities, one can assign nodes to communities first, and then place links based on that assignment. For a model of link communities where links are partitioned, however, one cannot assign links to communities until the links exist, so the links and their groupings have to be created at the same time. The key differences between Ball's model and our model, as explained below, hinges upon the above observation. Firstly, Ball's model is parameterized by a set of parameters $\theta_{ik}$'s, where $\theta_{ik}$ denotes the propensity of node $i$ to have links in the $k$-th community. These parameters, however, cannot precisely express the membership of node communities. On the other hand, our model is parameterized by a set of parameters $d_{ik}$'s, where $d_{ik}$ is defined as the expected node degree of $i$ in the $k$-th community. Thus, we can easily get the community membership of any node $i$ using the normalization of $d_{ik}$ according to (8). Secondly, Ball's model takes $\theta_{ik}\theta_{jk}$ as the expected number of links in the $k$-th community connecting nodes $i$ and $j$, which is a heuristic with little theoretical support. In contrast, our model takes $d_{ik}d_{jk}/\sum_s d_{sk}$ as the expected number of links in the $k$-th community between nodes $i$ and $j$, which is based on the widely accepted null model of modularity [5], and thus is statistically rigorous. More importantly, the main purpose of our model is not only to accommodate the coexistence of node and link communities, but also to support the hybrid node-link community scheme, which is our main contribution beyond the existing work focusing on node or link communities separately.

A practical issue in network structure analysis is the lack of the ground-truth or a reliable quality measure of a network. This issue is exacerbated on networks of overlapping structures since overlapping nodes often render ambiguous explanations. Most of the current quality measures are designed for non-overlapping structures. When extended, these methods penalize overlapping structures [36]. Fortunately, the generalized map equation [19] based on the principle of MDL and the community conductance [24] can naturally measure overlapping communities, which were the quality metrics used in this study.

Another critical issue is the lack of information of the number of communities to be targeted for. Neither robust criterion nor efficient method for this problem seems to be currently available [17]. A statistical method for model selection may in principle be able to find the number of communities, but it is at present too computationally demanding to be applicable to any but some small networks [13]. In our current study, we used two methods to determine the number of communities. First, we adopted the MDL as a yardstick to look for such network structures that can be encoded in minimum sizes. Second, for large networks, we devised a scheme of performing recursive bipartitioning until a terminal condition was met so that no number of communities needed to be determined *a priori*.

**ACKNOWLEDGEMENTS**

This work was supported by the National Science Foundation of USA (DBI-0743797), the National Institutes of Health of USA (R01GM100364), the National Natural Science Foundation of China (61133011, 61303110), and the China Scholarship Council (award to Dongxiao He).

**APPENDIX A. MAP EQUATION FOR OVERLAPPING COMMUNITIES**

The map equation for overlapping communities [19] measures how well we can compress a description of flow in the network when it is partitioned into communities with possible overlaps. The idea follows the principle of Minimum Description Length (MDL) that any regularity in the data can be used to compress the data [20]. If one can find a way to encode the path of random walk on the network and consider the overlapping community structure as the regularity in the network, the description length of path will be used to evaluate the quality of the overlapping communities.

In the map equation, the encoding rule for the path description can be described as follows. It uses the codebook at two levels: the first level code describes the communities with overlaps and the second level code distinguishes a specific node from other nodes in the same community. In this strategy, a community code (first level) should be recorded in the path description when the random walk enters a new community, and the random walks inside the community can be uniquely described by only recording the second level code. Besides, an exit code should be assigned to each community and it should be recorded when the random walk exits the community, so that the first level code and the second level codes can be distinguished.

Define a cover $M$ of a network $N$ as a set of communities such that each node is assigned to at least one community. The map equation $L(M)$ gives the average number of bits per step that it takes to describe an infinite random walk on the network with cover $M$:

$$L(M) = q_{out}H(Q) + \sum_{k=1}^{c} p_{in}^k H(P^k), \tag{A.1}$$

where $k$ is the index of community (note that we also use $k$ denote the $k$-th community for simplification), $i$ the index of node, and $c$ the number of communities; $q_{out} = \sum_{k=1}^{c} q_{out}^k$ is the total probability of using the first level codebook where $q_{out}^k$ is the probability of using the first level code for community $k$; $p_{in}^k = q_{out}^k + \sum_{i \in k} p_i^k$ is the probability of using the second level codebook and the exit code for community $k$, and $p_i^k$ is the probability of node $i$ being visited as a number of community $k$, which is equal to the probability of using the second level code for node $i$ in community $k$. $H(Q)$ is the average description length of the first level codebook:

$$H(Q) = -\sum_{k=1}^{c} \left( \frac{q_{out}^k}{q_{out}} \log \frac{q_{out}^k}{q_{out}} \right), \tag{A.2}$$

while $H(P^k)$ is the description length of the second level codebook for community $k$:

$$H(P^k) = -\frac{q_{out}^k}{p_{in}^k}\log\frac{q_{out}^k}{p_{in}^k} - \sum_{i\in k}\left(\frac{p_i^k}{p_{in}^k}\log\frac{p_i^k}{p_{in}^k}\right). \quad (A.3)$$

In order to compute the map equation for overlapping communities, we need to calculate the visit rates $p_i^k$ for all communities $k \in M_i$ which a node $i$ is assigned to, and the exit probabilities $q_{out}^k$ of all communities. Here $M_i$ denotes the set of community indexes of some node $i$.

The movements between multiply assigned nodes and overlapping communities are straightforward. Whenever the random walk arrives at a node that is assigned to multiple communities, it remains in the same community if possible or switches to one of the other communities randomly otherwise. For example, assuming that the walk is at a node $j$ in community $s$, it remains in community $s$ when moving to node $i$ if node $i$ is assigned to community $s$, $s \in M_i$. But if node $i$ is not assigned to community $s$, $s \notin M_i$, it switches with equal probability $1/|M_i|$ to any of the communities to which node $i$ is assigned. If the transition function is defined as

$$\delta_{js\to ik} = \begin{cases} 1 & \text{if } s = k \\ \frac{1}{|M_i|} & \text{if } s \neq k \text{ and } s \notin M_i \\ 0 & \text{if } s \neq k \text{ and } s \in M_i \end{cases}, \quad (A.4)$$

the visit rates can then be written as

$$p_i^k = \sum_j \sum_{s \in M_j} \left(p_j^s u_{ji} \delta_{js \to ik}\right), \quad (A.5)$$

where $u_{ji} = w_{ji}/\sum_r w_{jr}$ denotes the probability of the random walk moving from nodes $j$ to $i$. The visit rates $p_i^k$ can be computed with the fast iterative algorithm BICGSTAB [37]. Then since every node in community $k$ guides a fraction $\sum_{j \notin k} u_{ij}$ of its conditional probability $p_i^k$ to nodes outside community $k$, the exit probability of community $k$ is

$$q_{out}^k = \sum_{i \in k}\left(p_i^k \sum_{j \notin k} u_{ij}\right). \quad (A.6)$$

Using (A.5) and (A.6), the map equation in (A.1) can be derived.

**REFERENCES**
[1] A. Lancichinetti, S. Fortunato, Community detection algorithms: A comparative analysis, Phys. Rev. E 80(5) (2009) 056117.
[2] S. Fortunato, Community detection in graphs, Phys. Rep. 486 (2010) (2010)75-174.
[3] M. Coscia, F. Giannotti, D. Pedreschi, A classification for community discovery methods in complex networks, Stat. Anal. Data. Min. 4 (5) (2011) 512-546.
[4] M. Girvan, M. E. J. Newman, Community structure in social and biological networks, Proc. Natl.

Acad. Sci. USA 99 (12) (2002) 7821-7826.

[5] M. E. J. Newman, M. Girvan, Finding and evaluating community structure in networks, Phys. Rev. E 69 (2) (2004) 026113.

[6] V. D. Blondel, J. L. Guillaume, R. Lambiotte, E. Lefebvre, Fast unfolding of communities in large networks. J. Stat. Mech. 2008 (2008) P10008.

[7] M. Rosvall, C. T. Bergstrom, Maps of random walks on complex networks reveal community structure, Proc. Natl. Acad. Sci. USA 105 (4) (2008) 1118-1123.

[8] E. Y. K. Chan, D.-Y. Yeung, A convex formulation of modularity maximization for community detection, in: Proceedings of the Twenty-Second International Joint Conference on Artificial Intelligence (IJCAI), Barcelona, Spain, 2011, pp. 2218-2225.

[9] G. Palla, I. Derényi, I. Farkas, T. Vicsek, Uncovering the overlapping community structure of complex networks in nature and society, Nature 435 (7043) (2005) 814-818.

[10] Y.-Y. Ahn, J. P. Bagrow, S. Lehmann, Link communities reveal multiscale complexity in networks, Nature 466 (7307) (2010) 761-764.

[11] T. S. Evans, R. Lambiotte, Line graphs, link partitions, and overlapping communities, Phys. Rev. E 80 (1) (2009) 016105.

[12] Y. Kim, H. Jeong, Map equation for link communities, Phys. Rev. E 84 (2) (2011) 026110.

[13] B. Ball, B. Karrer, M. E. J. Newman, Efficient and principled method for detecting communities in networks, Phys. Rev. E 84 (3) (2011) 036103.

[14] D. He, D. Liu, W. Zhang, D. Jin, B. Yang, Discovering link communities in complex networks by exploiting link dynamics, J. Stat. Mech. 2012 (2012) P10015.

[15] D. E. Knuth, The Stanford GraphBase: A Platform for Combinatorial Computing, ACM Press, New York, 1994.

[16] M. E. J. Newman, S. H. Strogatz, D. J. Watts, Random graphs with arbitrary degree distributions and their applications, Phys. Rev. E 64 (2) (2001) 026118.

[17] B. Karrer, M. E. J. Newman, Stochastic blockmodels and community structure in networks, Phys. Rev. E 83 (1) (2011) 016107.

[18] A. P. Dempster, N. M. Laird, D. B. Rubin, Maximum-likelihood from incomplete data via the EM algorithm, J. R. Stat. Soc. Series B Stat. Methodol. 39 (1) (1977) 1-38.

[19] A. V. Esquivel, M. Rosvall, Compression of flow can reveal overlapping-module organization in networks, Phys. Rev. X 1 (2) (2011) 021025.

[20] P. D. Grünwald, The Minimum Description Length Principle, The MIT Press, Cambridge, MA, 2007.

[21] W. W. Zachary, An information flow model for conflict and fission in small groups, J. Anthropol. Res. 33 (4) (1977) 452-473.

[22] A. Lancichinetti, S. Fortunato, J. Kertész, Detecting the overlapping and hierarchical community structure in complex networks, New J. Phys. 11 (2009) (2009) 033015.

[23] The real-world networks we used here are available in http://www-personal.umich.edu/~mejn/netdata/. July, 10, 2013.

[24] J. Leskovec, K. J. Lang, M. W. Mahoney, Empirical comparison of algorithms for network community detection, in: Proceedings of the 19th International World Wide Web Conference (WWW), Raleigh, North Carolina, USA, 2010, pp. 631-640.

[25] I. Xenarios, D. W. Rice, L. Salwinski, M. K. Baron, E. M. Marcotte, D. Eisenberg, DIP: the Database of Interacting Proteins, Nucleic Acids Res. 28 (1) (2000) 289-291.

[26] M. Ashburner, C. A. Ball, J. A. Blake, D. Botstein, H. Butler, J. M. Cherry, A. P. Davis, K. Dolinski, S. S. Dwight, J. T. Eppig, M. A. Harris, D. P. Hill, L. Issel-Tarver, A. Kasarskis, S. Lewis, J. C. Matese, J. E. Richardson, M. Ringwald, G. M. Rubin, G. Sherlock, Gene ontology: tool for the unification of biology. The gene ontology consortium. Nat. Genet. 25 (1) (2000) 25-29.

[27] D. G. Altman, Practical Statistics for Medical Research, Chapman& Hall/CRC, London, 1991.

[28] D. L. Nelson, C. L. McEvoy, T. A. Schreiber, The University of South Florida, word association, rhyme, and word fragment norms, <http://w3.usf.edu/FreeAssociation/>. July, 10, 2013.

[29] C. Fellbaum, WordNet: An Electronical Lexical Database, The MIT Press, Cambridge, MA, 1998.

[30] M. E. J. Newman, E. A. Leicht, Mixture models and exploratory analysis in networks, Proc. Natl. Acad. Sci. USA 104 (23) (2007) 9564-9569.

[31] E. M. Airoldi, D. M. Blei, S. E. Fienberg, E. P. Xing, Mixed membership stochastic blockmodels, J. Mach. Learn. Res. 9 (2008) (2008) 1981-2014.

[32] W. Ren, G. Yan, X. Liao, L. Xiao, Simple probabilistic algorithm for detecting community structure, Phys. Rev. E 79 (3) (2009) 036111.

[33] H. Shen, X. Cheng, J. Guo, Exploring the structural regularities in networks, Phys. Rev. E 84 (5) (2011) 056111.

[34] K. Nowicki, T. A. B. Snijders, Estimation and prediction for stochastic blockstructures, J. Am. Stat. Assoc. 96 (455) (2001) 1077-1087.

[35] M. E. J. Newman, Communities, modules and large-scale structure in networks, Nat. Phys. 8 (1) (2012) 25-31.

[36] J. Xie, S. Kelley, B. K. Szymanski, Overlapping community detection in networks: the state of the art and comparative study, ACM Comput. Surveys 45 (4) (In press).

[37] H. A. van der Vorst, BI-CGSTAB: a fast and smoothly converging variant of BI-CG for the solution of nonsymmetric linear systems, SIAM J. Sci. Comput. 13 (2) (1992) 631-644.